\documentclass[sigconf]{acmart}

\renewcommand\footnotetextcopyrightpermission[1]{}
\acmConference[CHIWork '26]{CHIWork '26 Workshop: Interrogating GenAI Augmentation for CHIworkers}{June 22, 2026}{Linz, Austria}
\acmYear{2026}

\newcommand{\workshopNote}{%
  \begin{quote}
    \small\itshape
    This position paper is submitted to the CHIWork 2026 Workshop:
    \emph{Interrogating GenAI Augmentation for CHIworkers:
    Strategies for Professional Autonomy and Accountability}
    (June 22, 2026, Linz, Austria).
    Workshop proposal: \cite{sandhaus2026interrogating}.
  \end{quote}%
}

\title{Human-Centered Design: The Disclosure of
Generative Artificial Intelligence for Emerging Professionals}

\author{Sydney Lee}
\affiliation{%
  \institution{The University of North Carolina at Charlotte}
  \city{Charlotte}
  \country{USA}}
\email{alee164@charlotte.edu}

\renewcommand{\shortauthors}{Lee}

\begin{abstract}
  As the Human-centered Design continues to grow, generative AI has the potential to streamline the research process by iterating tasks within established workflows to increase efficiency. However, integrating AI raises concerns surrounding ethical bias, complexity, and the lack of prioritization of humanistic values. Emerging professionals represent a cohort with the opportunity to learn Human‑Centered Design principles, yet without this foundation AI becomes more of a crutch than a tool, leading to reduced experience with deep‑work, decreased autonomy, and deskilling of key foundations. Disclosures are a common method to self‑report AI usage, but they provide little clarification on appropriate implementation and may encourage omission to avoid consequences. This paper reflects on experiences in the Human‑Centered Design course ITIS8300, which emphasized optimizing user experience, enhancing innovation and collaboration, and improving efficiency through iterative user feedback. A semester‑long project, structured through milestones and team roles including a generative AI advocate, resulted in a high‑level disclosure report detailing design processes, methodology, findings, and rationale for AI usage. The course offered freedom in execution while setting clear boundaries for incorporating human feedback, reinforcing justification for HCI workflows and encouraging transparent AI use. This approach mirrors an industry with minimal regulation, demonstrating that when AI usage is safe, justified, and transparent, it can significantly advance the field through AI‑augmented workflows and support co‑creation an increase productivity.
  
\end{abstract}


\begin{document}
\maketitle

\workshopNote

\section{Introduction}

Generative AI has the potential to streamline the research process by iterating tasks within an established workflow to help researchers complete work more efficiently. However, integrating AI into these workflows raises concerns surrounding ethical bias, complexity, and the lack of prioritization of humanistic values. Emerging professionals are a unique cohort with the opportunity to learn Human‑Centered Design principles and apply them to everyday workflows while utilizing AI. However, without a strong foundation in these concepts, AI can become more of a crutch than a tool. This may leave new practitioners with limited experience in deep work, decreased autonomy, and the deskilling of research personnel. AI disclosures such as noting where AI was used in code or submitting a separate disclosure statement are intended to support transparency, yet they provide minimal clarification on appropriate implementation in specific contexts. This can lead to incomplete or omitted reporting, especially when emerging professionals fear negative consequences for utilizing AI in any capacity.

This paper examines these challenges by reflecting on the experiences in the Human‑Centered Design course (ITIS8300), which highlighted design methods through activities, lectures, and a semester long project. The course emphasized the strengths of traditional human‑ centered design, including optimizing user experience, enhancing innovation and collaboration, and improving cost and efficiency through iterative user feedback. The project was divided into milestones, team members were assigned a role including project manager, UX designer, and generative AI advocate to develop a transparency card for an on‑demand platform. Our reports documented the design process, meetings, methodology, findings, future implications, and rationale for AI usage, detailing opportunities to use AI, actual usage, and the reasoning behind our decisions.
The course left AI usage to the groups degression while setting clear boundaries for incorporating human feedback through usability testing and heuristic evaluations. This  required detailed justification for our workflows and offered feedback on AI's implementation at each milestone. These experiences reflects the current industry, in which companies set their own regulations for AI use. Given AI usage is safe, justified, and communicated transparently, it has the potential to advance the field through AI‑augmented workflows is substantial. This approach supports co‑creation with AI to increase productivity in researcher workflows.

\section{Background and Motivation}

Human‑centered design explores how researchers understand and reflect the lived experiences of users in research to accurately address their needs. Early literature highlights technological change, along with how style and quality impacts HCI to better define what interaction means\cite{10.1145/3325285}. This established the foundation to further explore perception, providing insight on users’ sensory, memory, and motor control and how these factors define a given users interface experience. This can improve design yet be misused, through raising concerns regarding ethics and user vulnerability. As design methods develop, contextual design introduce users as partners whose lived realities expose developer assumptions\cite{10.1145/97243.97304}. Thus, emphasizing that design must come before development to ensure technologies aligned with human needs.

Through exploring heuristic evaluation, qualitative, and quantitative methods, it is understood that user perspectives vary based on lived experience and therefore their performance in usability studies\cite{10.1145/97243.97281}. Additionally, there are limitations that come with evaluation methods including: a) subjectivity when conducting qualitative methods and b) the statistical results of quantitative methods tend to have stronger validity but miss more nuanced user challenges. As Human-centered design research continues to develop, industry is implementing design flows yet yields tension between user needs and stakeholder demands. This raises questions surrounding transparency, feature optimization, and prioritizing commercialization for profit\cite{chilana2015}. Overall, this literature depicts Human‑centered design as a way to understand users biologically, cognitively, and socially. The involvement of new technologies (e.g. Generative Artificial Intelligence) bring into question the efficiency of design methods, inclusion of industry and the constraints that come with it, and how technologies are impacting users experience.

\section{Position / Argument}

Understanding the strengths and weaknesses of integrating Generative AI into traditional Human‑Centered Design (HCD) necessitates a foundation in HCD concepts prior to AI use. The Human‑Centered Design course (ITIS8300) provided this context through lectures, academic papers, discussions, and interactive activities, along with a semester‑long project divided into milestones. The project focused on iteratively developing a transparency card for an on‑demand platform, allowing students to apply core HCD skills such as creating personas, building storyboards, conducting qualitative research, and prototyping. My group selected Expedia and examined how improving the existing transparency card and features could enhance user experience. This structure supported learning the research life cycle and meeting ACM peer‑review standards through submitting milestone reports.

Traditional HCD relied heavily on laboratory findings \cite{10.1145/97243.97304}. While this addressed certain design concerns, it often misapplied human factors by failing to reflect how users’ perceptions are shaped by their senses and past experiences. Contextual design, discussed in \cite{10.1145/97243.97304}, responded to these concerns by bringing users into the design process. Ultimately shifting power from developers to users by ensuring their mental models aligned with the technology they engage with prior to development. This reflects the importance of HCD which is to embed users’ expectations, experiences, and requirements into each step of the design process.

Within the course, these principles were reinforced through usability testing, heuristic evaluations, and iterative feedback. Our final disclosure report documented the design process, methodology, findings, and rationale for AI usage. This included opportunities to use AI, actual usage, and reasons behind our decisions. These observations show how structured, transparent practices can support integrating AI into HCD workflows while prioritizing user needs. 

\subsection{Human-Centered Design Limitations}
Obstacles presented by traditional human‑centered design methods involve the significant time required for user surveys, interviews, prototyping, and collecting iterative feedback. Time demands also arise during the developmental stages. For example, in Milestone 3, developing persona storyboards was an extensive process that required time not only to create the storyboards whether on paper or through Figma but also to identify explicit and implicit user feedback and consolidate it into design features that could improve the Expedia platform. We then collaborated as a team to determine which features were common, which differed, and which needed to be prioritized based on individual interviews. Additionally, in Milestone 2, we developed diagrams to establish themes across all interviews. This required substantial consolidation of user responses and instituted discussions within our team prior to writing our report. Although these milestones were completed within the allotted time, it is important to recognize that the project was scaled down for the course and took place over an entire semester. This demonstrates that a larger‑scale project with more resources, higher costs, and greater time commitments could place significant strain on a research team. Ensuring credibility, validity, and proper implementation of user feedback at all stages becomes increasingly challenging as the scope expands.

\subsection{Responsible AI Use}
This introduction of generative AI, has revealed benefits across multiple domains. These benefits range from civilians with no technical background being encouraged to learn more, to developers being able to streamline iterative tasks so they can work more efficiently and solve problems more effectively in industry. Researchers are also able to utilize generative AI through Google NotebookLM or citation‑management platforms that help organize resources across papers. As discussed in the course, from the 1960s to the early 1980s there was a strong emphasis on productivity, especially in \cite{10.1145/3325285} CHI proceedings establishing the definition of interaction and how it supports productivity for researchers and the technological field as a whole. The rapid development and implementation of generative AI, emphasizes that this age of productivity is still very much alive. However, even as we use technology to streamline work processes, it is important to understand that human‑in‑the‑loop feedback is essential. This ensures that over‑reliance, cognitive dependence, and ethical biases are not embedded into the technology we develop and keeping the focus on users and humanistic values to ensure equitable innovation that improves accessibility and technology for all.

As a solution, AI can address several issues within traditional human‑centered design. One major area is time consumption, which generative AI can help reduce. For example, in Milestone 5, AI could have supported documentation by correcting grammar and sentence structure, reducing the pressure of developing a cohesive and thorough report that expanded on Milestones 1 through 4. Consolidating information is not inherently difficult, but the one‑week time constraint added pressure to ensure user feedback was properly implemented into a high‑fidelity prototype while also completing the report itself. AI could have streamlined this process by highlighting key areas from earlier milestones that were essential to include in Milestone 5, along with integrating evaluation feedback. User evaluation feedback can be subjective and abstract, since users’ interactions with the interface were considered from multiple perspectives driving from behavioral, verbal, and even facial expressions. Consolidating and interpreting this feedback took a large portion of our time to ensure proper development of features into the high‑fidelity prototype. AI could have helped streamline this interpretation process so that once evaluations were processed, we could identify key areas needing improvement and devote more time to developing the report and refining the high‑fidelity prototype features implemented in Figma.

\subsection{Irresponsible AI Use}
These are the issues that are presented through utilizing AI. This solution creates obstacles regarding ethical bias, added complexity, and offloading control over design which hinders skill development. Provided with clear metrics and technical responsibility, AI can produce robust and well‑completed solutions that meet the needs of a large population. However, when subjectivity and creativity must be considered or embedded within a prototype or system, AI can introduce bias that challenges the credibility and ethical considerations of a research project. The value of generative AI is not in question, but it must be used in ways that leverage its strengths and highlight clear areas of improvement. This prevents it from becoming a crutch that encourages users to offload cognitive input, visual design, or developmental skills related to Figma or research dissemination.

AI can also add unnecessary complexity to a system. For example, in Milestones 1B and 2, we attempted to use generative AI to streamline filtering options, which were a significant struggle for users experiencing decision paralysis. Many users mentioned they would leverage AI to mitigate this issue, and a large majority already use AI to find better pricing and travel options rather than relying on on‑demand platforms like Expedia. However, creating a transparency card for this feature was extremely difficult. A large portion of the interface’s interactivity was being offloaded to generative AI, making it unclear how to strategically place improved features. We were unsure how to improve Expedia’s transparency in a way that would allow generative AI to iterate through the platform effectively and return accurate information. We also did not know whether we should provide a transparency card for generative AI tools like ChatGPT, which do not include certain features users desired, but Expedia includes.

Additionally, we were unsure what narrative or perspective we, as technical developers or researchers, were imposing on users by taking the approach of using generative AI for booking. Stakeholders who were interested in using AI and those who had experience streamlining their process with AI varied widely. This made it difficult to identify a target demographic that reflected both interested and experienced users for this pathway. It also made it challenging to avoid leading or suggestive questions based on users’ interest in AI, rather than exploring whether this was an approach they naturally used or considered on their own.

 While generative AI did not seem to be an efficient way forward within the constraints of our project due to the nature of the course and our negative early experiences this does not mean the tool cannot be implemented in practitioner workflows. However, we must be considerate of how we use it alongside user information and data‑collection methods. We must also ensure that we are not offloading the essential development of new skills as researchers for the sake of streamlining complex processes or being cost‑efficient. Productivity should not surpass the need for human input, creativity, and knowledge development. Instead, these values must rise to meet the unique demands of productivity in this new age.

 \subsection{Proposed Practice}
 Ultimately, utilizing AI in strategic ways throughout the research process whether through defining the research topic, identifying gaps, or establishing user testing should position AI as a co‑creator rather than offloading responsibilities entirely. This prevents researchers from relinquishing complete control. This can be achieved by communicating the workflow the research team will follow, collaboratively deciding which areas AI will be leveraged in, and determining what type of AI will be used. Detailing these processes ensures the entire team is on the same page and that AI usage is strategically mapped out, defined, and defensible within the research report. This also decreases complexity or ambiguity, not only in how AI will be leveraged, but also in your ability to justify why it is being used and how it benefits the research outcome without jeopardizing ethical considerations.
Additionally, it is essential to keep human‑centered design principles at the forefront of all design and research implementation. This ensures transparency in every decision, trust through the validity and credibility of the research, and a continued emphasis on the human‑in‑the‑loop approach so that humanistic values remain present at all points of the system.
Finally, ensuring that AI is leveraged for low‑stakes but high‑impact work such as automated tasks like labeling data after it has been collected can help the research team offload work without jeopardizing humanistic problem‑solving or introducing bias. Through utilizing these strategies, we can decentralize the implementation of AI within traditional human‑centered design principles. This enhances the human‑centered approach through ensuring that a natural human experience is reflected in research. As a result, we are able to extrapolate rich and accurate data that more effectively contributes knowledge to the field.

\section{Implications and Discussion Points}

Practitioners should utilize AI in two key ways. The first is to ensure that the system is being used in ways that align with its specific strengths. For example, ChatGPT, Gemini, and Claude are all different models with different objectives. Some are more sycophantic, while others are meant to aid in technical workflows. Understanding the system you are working with and its strengths and weaknesses creates clear boundaries for not only where it is used within your project process, but also how you use it.
Additionally, researchers must develop the necessary skills to discern where bias is being imbued and where ethical concerns arise when using these models. It is imperative that we do not offload the essential educational skills that are developed through working through the human‑centered design process. Understanding the foundations of tools such as Figma for designing and implementing requirement features, being able to conduct research dissemination and develop a peer‑reviewed‑level ACM report, and interpreting user behavioral responses and verbal exclamations while they engage with a platform are all values that must be upheld. These are necessary skills that define how researchers make progress within a field and grow to contribute knowledge beyond what a technical system can provide.
As developing researchers, it is essential to understand project design flow and also remain personable with users. We must develop the skills needed to engage with them not only for the purpose of the study, but also to ensure that we accurately reflect, interpret, and implement their lived experiences in a way that remains true to those experiences and accurately represents the human experience. This includes encapsulating morals, values, perceptions, and understandings that help shed light on future research or work that builds on our own. This is possible when AI is implemented after these necessary research skills have been developed.

\subsection{Accountability Implications}
This paper advocated for the responsible integration of accountability to guide the fusion of human-centered design principals and generative AI through three ways: a) accountability through transparency, b) intentional usage and c) responsible system use. These practices help identify the misalignment that may hinder or negatively impact users. Accountability provides a feasible way to identify and address these challenges at the source. Generative AI is not inherently bad, but when misused this can hinder the efficiency and productivity of work. This tool, can become a crutch if used without intention effectively substituting human responsibility and posing a threat to ethical collection of user data. This course established these three areas of accountability through requiring documentation of generative AI usage and assigning a 'Gen AI Advocate' role for the project(all students within the group alternated roles). This advocate is responsible for explaining the opportunities where AI could have been used, the rational backing, actual usage, and justification for these decisions. This process highlighted the importance of understanding the ethical risks as well as the benefits of AI knowledge acquisition and project progress. 

Additionally, students were able to exercise human-centered design skills through weekly engagement with course content, discussion posts, quizzes, and in-class activities that evaluate our understanding. Throughout the project life cycle skills such as interpreting user behaviors and lived experience in order to produce a graduate level report. 

This course offered a unique opportunity to develop Human centered design skills because this class was not siloed into a single discipline. Instead of separating research methods, prototyping, and research processes, this course integrated all three. Thus, allowing students to leverage each concept for the development of a research project that prepares students for industry level work while contributing effectively to the field of Computer Science. 

\begin{acks}
  \textbf{AI Use Statement:}
  
During the development of this report Grammerly and Copilot were utilized for editing of gammer and sentence structure. All brainstorming, drafting, claims, experiences, analysis, and framing were my own with revision suggestions from my advisor. All AI-generated text has been reviewed and revised for the submission of this position paper.

\end{acks}

\bibliographystyle{ACM-Reference-Format}
\bibliography{references}

\end{document}